\documentclass[9pt,twocolumn,twoside]{osajnl}

\usepackage[T1]{fontenc}

\journal{josab}
\setboolean{shortarticle}{false}

\usepackage{amssymb}
\usepackage{amsmath}
\usepackage{graphicx}
\usepackage{array}
\usepackage{multirow}
\usepackage{widetext}
\usepackage{color}

\newcommand{\pare}[1]{\left( #1 \right)}

\newcommand{\llav}[1]{\left\lbrace #1 \right\rbrace}

    \title{Virtual-State Spectroscopy with Frequency-Tailored Intense Entangled
    Beams}

\author[1,2,*]{J. Svozil\'{i}k}

\affil[1]{Quantum  Optics  Laboratory,  Universidad  de  los  Andes,  A.A.
    4976,  Bogot\'a  D.C.,  Colombia}

\author[2]{J. Pe\v{r}ina Jr.}
\affil[2]{Joint Laboratory of Optics of Palack\'y University and  Institute of 
Physics of CAS, Faculty of Science, Palack\'y University, 17. listopadu 12, 771 
46 Olomouc, Czech Republic}

\author[3]{R. de J. Le\'{o}n-Montiel}
\affil[3]{Instituto de Ciencias Nucleares, Universidad Nacional
    Aut\'{o}noma de M\'{e}xico, Apartado Postal 70-543, 04510 Cd. Mx.,
    M\'{e}xico}

\affil[*]{jiri.svozilik@gmail.com}

\begin{abstract}

    In this contribution we analyze virtual-state spectroscopy --- a unique
    tool for extracting information about the virtual states that
    contribute to the two-photon excitation of an absorbing medium
    --- as implemented by means of intense entangled beams with tunable
    spectral correlations. We provide a thorough description of all
    contributing terms (classical and quantum) in the two-photon
    absorption signal, as well as the limits imposed by the power of
    the pump that produces the entangled beams on the observability of
    the spectral lines of the virtual transitions. We find that
    virtual-state spectroscopy may be implemented with entangled
    twin beams carrying up to $10^4$ photon pairs. This implies that, in
    principle, one might be able to detect two-photon absorption
    signals up to four orders of magnitude larger than previously
    reported, thus paving the way towards the first experimental
    realization of the virtual-state spectroscopy technique.
\end{abstract}

\ociscodes{(300.6420) Spectroscopy, nonlinear; (020.1335) Atom optics}

\begin{document}

\pagestyle{plain}
   \maketitle
\section{Introduction}
Nonlinear spectroscopy has become an invaluable tool across many
fields of research
\cite{Shen1984Principles,Mukamel1999Principles}. In particular,
two-photon absorption (TPA) spectroscopy [for the scheme, see
Fig.~1] has allowed us to obtain information about a sample that
would not be accessible otherwise. Interestingly, the use of
entangled light in two-photon spectroscopy has received a great
deal of attention very recently
\cite{dayan2004two,kojima2004,Lee2006Entangled,Roslyak2009,guzman2010,oka2010,oka2011,oka2011-2,salazar2012,Schlawin2012,Raymer2013,kalashnikov2014,dorfman2016,kalashnikov2016,schlawin2017,villabona2017,varnavski2017,schlawin2017_njp}
because of the unique phenomena that arise in the interaction of
entangled photon pairs with matter. Examples of these effects are
the linear scaling of the TPA rates on the photon flux
\cite{Javanainen1990}, two-photon-induced transparency
\cite{Fei1997Entanglement}, the ability to select different states
in complex biological aggregates \cite{Schlawin2013Suppression},
and the control of entanglement in matter
\cite{Shapiro2011,Shapiro_book}.  Indeed, the prediction and
observation of these fascinating effects can be understood as a
direct consequence of the dependence of the TPA signal on the
properties of quantum light that interacts with the sample
\cite{Perina1991Quantum}.

Among different techniques proposed over the years,
entangled-photon virtual-state spectroscopy (VSS)
\cite{Saleh1998Entangled,Perina1998Multiphoton,Leon2013Role} has
proved to be a unique tool for extracting information about the
virtual states --- energy non-conserving atomic transitions
\cite{Shore1979,SakuraiBook} --- that contribute to the two-photon
excitation of an absorbing medium. In this technique,
virtual-state transitions, a signature of the medium, are
experimentally revealed by introducing a time-delay between
frequency-correlated photons, and averaging over experimental
realizations differing in temporal correlations between them
\cite{Saleh1998Entangled}.

One important aspect of VSS is that the overall TPA rate, $R$, can be expressed as \cite{Fei1997Entanglement}
\begin{equation}
R = \sigma_{\rm E}\phi + \delta_{\rm r}\phi^{2},
\end{equation}
where $\sigma_{\rm E}$ is the entangled-light absorption cross
section, $\delta_{\rm r}$ is the random (classical) TPA cross section,
and $\phi$ is the flux density of photon pairs. Notice that Eq.
(1) states that entangled-photon effects will dominate the TPA
signal only when the photon-flux density is sufficiently small.
This conclusion has two opposite points of view when discussing
the experimental implementation of VSS. On one hand, the low
efficiency of spontaneous down-conversion in nonlinear crystals
guarantees the low photon-flux condition but, on the other, a low
photon-flux might result in an extremely weak TPA signal that
might require long measuring times, thus making the implementation
of virtual-state spectroscopy an unrealistic endeavor.

With the advent of ultrahigh flux sources of entangled photons
\cite{Brambilla2004,Jedrkiewicz2004,Bondani2007,Blanchet2008,dayan2005,shimizu2009},
one naturally wonders whether VSS can benefit from them,
especially because it has been shown that strong frequency
correlations between entangled beams persist at high photon-flux
conditions
\cite{Schlawin2013Photon,PerinaJr2015a,PerinaJr2016,PerinaJr2016a}.
Consequently, in this paper, we provide a thorough analysis of VSS
when implemented with intense entangled fields (further twin
beams). This allows us to determine the entangled-photon flux
threshold at which spectroscopic information about an absorbing
medium can be retrieved. Surprisingly, we find that virtual-state
spectroscopy may be implemented with twin beams carrying up to
$10^4$ photon pairs. This means that the strength of typical TPA
signals could be enhanced by up to four orders of magnitude, thus
paving the way towards the first experimental realization of
virtual-state spectroscopy.

The paper is structured as follows. In Sec. II we describe the
generation of twin beams, produced in a nonlinear crystal pumped
by an intense laser pulse. In Sec. III, we derive the explicit
form of the TPA signal for temporally-delayed intense twin beams.
In Sec. IV, we provide an example of the VSS implementation with
intense twin beams by extracting the energy level structure of a
model system, whose two-photon excitation takes place via three
intermediate states with randomly chosen energies. Finally, in
Sec. V we present our conclusions.

\begin{figure}
    \centering
    \includegraphics[width=8.5cm]{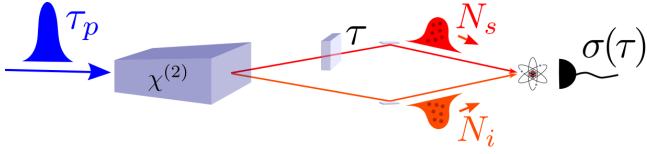}
    \caption{Schematic representation of the two-photon absorption process caused by two mutually-delayed
    beams that form a common twin beam. A nonlinear $\chi^{(2)}$ crystal is pumped by an intense laser pulse, thus producing two intense entangled beams.
    These beams interact with an absorbing medium and the two-photon absorption signal is measured as a function of an externally-introduced delay $\tau$
    between them; $\tau_{\mathrm p}$ stands for the pump-pulse duration and
    $N_{\mathrm s,i}$ is the number of photons present in the signal and
    idler beams, respectively.}
\end{figure}

\section{Generation of intense twin beams}
To describe the generation of twin beams, we follow the procedure
used by previous authors in Refs.
\cite{Wasilewski2006Pulsed,McKinstrie2009, McKinstrie2013,
christ2011,christ2013,Schlawin2013Photon,PerinaJr2015a}. The first
step is to solve the Schr\"{o}dinger equation in the first-order
perturbation approximation to obtain the two-photon spectral
amplitude. In the second step the Schmidt decomposition of the
two-photon spectral amplitude (TSA) of the generated photons is
calculated. This provides the Schmidt modes of the correlated
photon fields. The twin-beam description is then found by solving
the Heisenberg equations for the creation and annihilation
operators of the corresponding Schmidt spectral modes.

Let us start with the description of Spontaneous Parametric
Down-conversion (SPDC) in the Schr\"{o}dinger picture, based on the
momentum operator
\begin{eqnarray}
 \hat{G}(z) &=& 4\epsilon_{\mathrm 0}A\int_{-\infty}^{\infty}dt\chi^{(2)}
  \mathcal{E}_{\mathrm p}^{(+)}(z,t)\hat{E}_{\mathrm
  s}^{(-)}(z,t)\hat{E}_{\mathrm
  i}^{(-)}(z,t) \nonumber \\
 & &  +{\rm h.c.}
\label{Eq:1}
\end{eqnarray}
where $\epsilon_\mathrm{0}$ is the vacuum permittivity, $A$ is the traverse
area of interaction, $\chi^{(2)}$ is the second-order non-linear
susceptibility and h.c. stands for the Hermitian-conjugated term.
The pump field is modeled as a classical non-depleted field with
the form
\begin{equation}
\mathcal{E}_\mathrm{p}^{(+)}(z,t)=\frac{1}{\sqrt{2\pi}}\int_{-\infty}^\infty
d\omega_\mathrm{p} \; e_\mathrm{p}\left(\omega_\mathrm{p}\right)
\exp[-i(\omega_\mathrm{p} t-k_\mathrm{zp}z)]. \label{Eq:2}
\end{equation}
Notice that the spatial dependence is encapsulated in the
longitudinal wave vector $k_\mathrm{zp}$. Transversally, we consider the
pump homogeneous in the whole area $A$. The pump spectrum is given
by
\begin{equation}
 e_\mathrm{p}\left(\omega_\mathrm{p}\right)=\xi_\mathrm{p}\sqrt{\frac{\tau_\mathrm{p}}{\sqrt{2\pi}}}
 \exp\left[-\frac{\tau^2_\mathrm{p}}{4}\left(\omega_\mathrm{p}-\omega^0_\mathrm{p}\right)^2\right].
 \label{Eq:3}
\end{equation}
Here, the pump pulse duration is denoted by $\tau_{\mathrm p}$ and the pump
amplitude is described by
$\xi_\mathrm{p}=\sqrt{P_\mathrm{p}/\epsilon_{0}cfn_\mathrm{p}}$,
which depends on the pump power $P_\mathrm{p}$ and the repetition frequency
$f$; $n_\mathrm{p}$ is the index of refraction of the crystal at the pump
wavelength, and $c$ is the speed of light.  Finally, the generated
signal and idler photons are treated in a quantum manner and they
are described by the positive-frequency electric-field operator
amplitude
\begin{eqnarray}
 \hat{E}^{(+)}_{\mathrm
 j}(z,t)&=&\frac{i}{\sqrt{2\pi}}\int_{-\infty}^{\infty}d\omega_\mathrm{j}
 \sqrt{\frac{\hbar\omega_\mathrm{j}}{2\epsilon_\mathrm{0}Acn_{\mathrm
 j}}}\hat{a}\left(\omega_\mathrm{j}\right)
  \nonumber \\
 & & \mbox{} \times \exp[-i(\omega_\mathrm{j}t-k_\mathrm{j}z)],
\label{Eq:4}
\end{eqnarray}
with $\mathrm{j=s,i}$; $\hbar$ is the reduced Planck constant and
$\hat{a}_\mathrm{j}(\omega_\mathrm{j})$ is the bosonic annihilation operator of
the $\mathrm{j}$th photon with frequency $\omega_{j}$.

The first-order perturbation solution of the Schr\"{o}dinger
equation results in the following entangled two-photon state
\begin{eqnarray}
|\Phi\rangle&=&-\frac{i}{\hbar}\int_{0}^{L}dz\hat{G}\left(z\right)|{\rm
vac}\rangle=
\int_{-\infty}^{\infty}d\omega_\mathrm{s}\int_{-\infty}^{\infty}d\omega_\mathrm{i}\nonumber\\
 &
 &\Phi\left(\omega_\mathrm{s},\omega_\mathrm{i}\right)\hat{a}^{\dagger}_{\mathrm{s}}
 \left(\omega_\mathrm{s}\right)\hat{a}^{\dagger}_{\mathrm
 i}\left(\omega_\mathrm{i}\right)|{\rm vac}\rangle,
\label{Eq:5}
\end{eqnarray}
where $L$ stands for the length of the non-linear crystal and $
|{\rm vac}\rangle $ denotes the vacuum state of the signal and
idler fields in front of the crystal. The TSA
$\Phi\left(\omega_{\mathrm s},\omega_{\mathrm i}\right)$ is given
by
\begin{eqnarray}
 \Phi\left(\omega_{\mathrm
 s},\omega_\mathrm{i}\right)&=&\frac{i\chi^{(2)}\xi_{\mathrm p}L}{\sqrt{2\pi
  n_{\mathrm p}n_{\mathrm s}n_{\mathrm
  i}}}\sqrt{\frac{\tau_\mathrm{p}}{\sqrt{2\pi}}}\mathrm{sinc}\left[\Delta
k_{\mathrm{z}}\left(\omega_{\mathrm s},\omega_{\mathrm
i}\right)\frac{L}{2}\right]
  \nonumber \\
 & &  \hspace{-15mm} \mbox{} \times \exp\left[-\frac{\tau^2_\mathrm
 {p}}{4}\left(\omega_{\mathrm s}+\omega_{\mathrm
 i}-\omega^{0}_{\mathrm p}\right)^2-i\Delta k_{\mathrm{z}}\left(\omega_{\mathrm
 s},\omega_{\mathrm
  i}\right)\frac{L}{2}\right]
\label{Eq:6}
\end{eqnarray}
with $\Delta k_{\mathrm{z}}\left(\omega_{\mathrm
s},\omega_{\mathrm i}\right)=k_{\mathrm zp}\left(\omega_{\mathrm
s}+\omega_{\mathrm i}\right)-k_{\mathrm
zs}\left(\omega_\mathrm{s}\right)-k_{\mathrm
zi}\left(\omega_{\mathrm i}\right)$.

We now proceed with the second step of the calculation, where the
Schmidt decomposition is applied to find pairs of spectral modes
\cite{Law2000Continous,Law2004Analysis} of the normalized TSA,
that is,
\begin{equation}
\tilde{\Phi}\left(\omega_{\mathrm s},\omega_{\mathrm
i}\right)=\sum_{g=1}^{\infty}\lambda_{g}
f^{*}_{\mathrm{s},g}\left(\omega_{\mathrm
s}\right)f^{*}_{\mathrm{i},g}\left(\omega_{\mathrm i}\right),
\label{Eq:8}
\end{equation}
with $\Phi=\mathcal{N}L\tilde{\Phi}$ and $\mathcal{N}^2 L^2=\int
d\omega_{\mathrm s}\int d\omega_{\mathrm
i}|\Phi\left(\omega_{\mathrm s},\omega_{\mathrm i}\right)|^2$,
where $\mathcal{N}$ is a numerically-calculated normalization
constant. Notice that in Eq.~(\ref{Eq:8}),
$\left\{\lambda_g\right\}_{g=1}^{\infty}$ describes the set of
eigenvalues with corresponding eigenfunctions
$\left\{f_{\mathrm{s},g}(\omega_{\mathrm
s})\right\}_{g=1}^{\infty}$ and
$\left\{f_{\mathrm{i},g}(\omega_{\mathrm
i})\right\}_{g=1}^{\infty}$.

Thus, by making use of this formalism, we can rewrite the state in
Eq.~(\ref{Eq:5}) as
\begin{eqnarray}
|\Phi\rangle&=&\mathcal{N}L\sum_{g=1}^{\infty}\lambda_{g}\int
d\omega_{\mathrm s}\int d\omega_{\mathrm i}
f^{*}_{\mathrm{s},g}\left(\omega_{\mathrm
s}\right)f^{*}_{\mathrm{i},g}\left(\omega_{\mathrm
i}\right)\hat{a}^{\dagger}_{\mathrm{s}}\hat{a}^{\dagger}_{\mathrm{i}}|{\rm
vac}\rangle \nonumber\\
&=&\mathcal{N}\sum_{g=1}^{\infty}\lambda_g\hat{a}^{\dagger}_{\mathrm{s},g}\hat{a}^{\dagger}_{\mathrm{i},g}|{\rm
vac}\rangle. \label{Eq:9}
\end{eqnarray}
Note that the signal- (idler-) field creation operators
$\hat{a}^{\dagger}_{\mathrm{s},g}$ ($\hat{a}^{\dagger}_{\mathrm{i},g}$) of
independent Schmidt modes may be related to the spectral ones by
writing
\begin{equation}
\hat{a}_{\mathrm
s}\left(\omega_{\mathrm{s}}\right)=\sum_{g=1}^{\infty}f_{\mathrm{s},g}^{*}\left(\omega_{\mathrm{s}}\right)\hat{a}_{\mathrm{s},g},
\label{Eq:10}
\end{equation}
and similarly for the idler field.

Using the newly introduced operators of the Schmidt modes, the
operator $\hat{G}\left(z\right)$ can be rewritten as
\begin{equation}
\hat{G}\left(z\right)=i\hbar\mathcal{N}\sum_{g=1}^{\infty}{\lambda_g}\hat{a}^{\dagger}_{\mathrm{s},g}\hat{a}^{\dagger}_{\mathrm{i},g}
 + {\rm h.c.}
\label{Eq:11}
\end{equation}
Finally, by using this simplified form of the momentum operator,
we can find the evolution of the operators that describe the
produced twin beam, i.e. $\hat{a}^{\dagger}_{\mathrm{s},g}$ and
$\hat{a}^{\dagger}_{\mathrm{i},g}$. To do this, we use the Heisenberg
equations with the momentum operator $ \hat{G} $ given in
Eq.~(\ref{Eq:11}), that is,
\begin{eqnarray}
\frac{\partial \hat{a}_{\mathrm{s},g}}{\partial
z}&=&\frac{i}{\hbar}\left[\hat{G},\hat{a}_{\mathrm{s},g}\right]=\lambda_g\mathcal{N}\hat{a}^{\dagger}_{\mathrm{i},g},\nonumber\\
\frac{\partial \hat{a}_{\mathrm{i},g}}{\partial
z}&=&\frac{i}{\hbar}\left[\hat{G},\hat{a}_{\mathrm{i},g}\right]=\lambda_g\mathcal{N}\hat{a}^{\dagger}_{\mathrm{s},g}.
\end{eqnarray}
Their solution takes the form
\begin{eqnarray}
\hat{a}_{\mathrm{s},g}\left(L\right)&=&
u_g\hat{a}_{\mathrm{s},g}\left(0\right)+v_g\hat{a}^{\dagger}_{\mathrm{i},g}\left(0\right),\nonumber\\
\hat{a}_{\mathrm{i},g}\left(L\right)&=&
u_g\hat{a}_{\mathrm{i},g}\left(0\right)+v_g\hat{a}^{\dagger}_{\mathrm{s},g}\left(0\right),
\label{Eq:13}
\end{eqnarray}
where $u_g=\cosh(\mathcal{N}\lambda_g)$ and
$v_g=\sinh(\mathcal{N}\lambda_g)$. Notice from Eq.~(\ref{Eq:13})
that one can easily determine the number of signal (idler)
photons, contained in all modes, by writing
\begin{equation}
 N_\mathrm{s}=\int d\omega_{\mathrm
 s}\langle\hat{a}^{\dagger}_{\mathrm{s}}\left(\omega_{\mathrm
 s}\right)\hat{a}_{\mathrm{s}}\left(\omega_{\mathrm s}\right)\rangle
  =\sum_{g=1}^{\infty}\langle\hat{a}^{\dagger}_{\mathrm{s},g}\hat{a}_{\mathrm{s},g}\rangle=\sum_{g=1}^{\infty}|v_g|^2,
\end{equation}
whereas the effective amount of spectral modes, related to the
Schmidt number K, is obtained as
\begin{equation}
K_{\mathrm
UV}=\frac{\left(\sum_{g=1}^{\infty}u_gv_g\right)^2}{\sum_{g=1}^{\infty}
u_g^2v^2_g}.
\end{equation}

Before concluding this section, it is important to remark that
although the theoretical model described above is valid for most
experimental setups and applications, one most keep in mind that,
as any model, it has its limitations; particularly when extremely
high photon fluxes are considered
\cite{christ2013,Perez2014,PerinaJr2016a}.

\section{Interaction of twin beams with matter}

We now consider the interaction of twin beams with an absorbing
medium. For the sake of clarity, we assume a simple energy level
configuration of the medium where two-photon transitions occur
from a ground (initial) state $|\mathrm{g}\rangle$ to a
doubly-excited final state $|\mathrm{f}\rangle$ via non-resonant
intermediate states denoted by $|k\rangle$. For simplicity, we
omit any other degree of freedom connected to vibrational spectra
of the sample, and assume that the lifetimes of intermediate
states are longer than light-matter interaction time. This
approximation, which can be satisfied by selecting a proper
correlation time between photons
\cite{Fei1997Entanglement,Saleh1998Entangled,Leon2013Role,Schlawin2013Suppression},
implies that effects due to dissipation in the single-excitation
manifold (intermediate states) are assumed to be negligible.

The interaction of the electromagnetic field $\hat{E}$ and the
sample, in the dipole approximation, can be expressed as
\begin{equation}
\hat{H}(t)=-\hat{d}(t)\hat{E}(t)+ {\rm h.c.},
\end{equation}
where $\hat{d}$ is the dipole-moment operator, whose time
evolution is given by

	\begin{equation}
	\hat{d}(t)=\hat{\mu}_{kg}\exp[i(\varepsilon_{k}-\varepsilon_{\mathrm{g}})t]
	\end{equation}
	with $\hat{\mu}_{kg}$ being the single-excitation transition amplitude 
	operator from a state $|\mathrm{g}\rangle$ (with energy 
	$\varepsilon_{\mathrm{g}}$) to a state $|k\rangle$ (with energy 
	$\varepsilon_{k}$).

By considering that the medium is initially in its ground state
$|\mathrm{g}\rangle$, one can make use of second-order
time-dependent perturbation theory to find that the resulting TPA
signal is given by \cite{Perina1998Multiphoton,Leon2013Role}
\begin{eqnarray}\label{Eq:Probability1}
S_{\mathrm{g}\rightarrow
\mathrm{f}}&=&\frac{1}{\hbar^4}\int_{-\infty}^{\infty}dt_{\mathrm{2}}\int_{-\infty}^{t_{\mathrm{2}}}dt_{\mathrm{1}}\int_{-\infty}^{\infty}dt'_{\mathrm{2}}\int_{-\infty}^{t'_{\mathrm{2}}}dt'_{\mathrm{1}}
 M^{*}\left(t_{\mathrm{2}},t_{\mathrm{1}}\right)\nonumber\\
&\times&M\left(t'_{\mathrm{2}},t'_{\mathrm{1}}\right)\langle\hat{E}^{(-)}\left(t_{\mathrm{2}}\right)\hat{E}^{(-)}\left(t_{\mathrm{1}}\right)\hat{E}^{(+)}\left(t'_{\mathrm{2}}\right)\hat{E}^{(+)}\left(t'_{\mathrm{1}}\right)\rangle,
 \nonumber\\
\end{eqnarray}
where
$\langle\hat{E}^{(-)}\left(t_{\mathrm{2}}\right)\hat{E}^{(-)}\left(t_{\mathrm{1}}\right)\hat{E}^{(+)}\left(t'_{\mathrm{2}}\right)\hat{E}^{(+)}\left(t'_{\mathrm{1}}\right)\rangle$
corresponds to the four-point correlation function of the optical
field. The electric-field operator amplitude is defined as
\begin{equation}\label{Eq:Elec}
\hat{E}^{(+)}\left(t\right)=\hat{E}_\mathrm{s}^{(+)}\left(t\right)+\hat{E}_\mathrm{i}^{(+)}\left(t\right),
\end{equation}
and
\begin{eqnarray}\label{Eq:M}
 M(t_\mathrm{2},t_\mathrm{1})&=&\sum_{k}\mu_{\mathrm{f}k}\mu_{k\mathrm{g}}\exp\left[i\left(\varepsilon_{\mathrm{f}}-\varepsilon_{k}\right)t_\mathrm{2}
  \right. \nonumber \\
 & & \left. \mbox{}+i\left(\varepsilon_{k}-\varepsilon_{\mathrm{g}}\right)t_\mathrm{1}\right].
\end{eqnarray}
where $\mu_{fk}$ denotes the transition dipole moment from the $k$th 
intermediate state to the final doubly-excited state $|f\rangle$ and $\mu_{kg}$ 
is defined as in Eq.(17). Notice that the sum over the $k$ states in 
Eq.~(\ref{Eq:M}) appears because the excitation of the medium occurs through 
non-resonant intermediate states.

Upon substitution of Eq.~(\ref{Eq:Elec}) into
Eq.~(\ref{Eq:Probability1}), one finds that the TPA signal is
composed by sixteen different terms:

\begin{widetext}
\begin{eqnarray}\label{Eq:Probability2}
S_{\mathrm{g\rightarrow
f}}\left(\tau_\mathrm{s},\tau_\mathrm{i}\right)&=&\frac{1}{\hbar^4}\int_{-\infty}^{\infty}dt_{\mathrm{2}}\int_{-\infty}^{t_{\mathrm{2}}}dt_{\mathrm{1}}\int_{-\infty}^{\infty}dt'_{\mathrm{2}}\int_{-\infty}^{t'_{\mathrm{2}}}dt'_{\mathrm{1}}
 M^{*}\left(t_{\mathrm{2}},t_{\mathrm{1}}\right)M\left(t'_{\mathrm{2}},t'_{\mathrm{1}}\right)\nonumber\\
&&\hspace{3mm}\times\bigg[\langle\hat{E}_{\mathrm{s}}^{(-)}\left(t_{\mathrm{2}}\right)\hat{E}_{\mathrm{s}}^{(-)}\left(t_{\mathrm{1}}\right)\hat{E}_{\mathrm{s}}^{(+)}\left(t'_{\mathrm{2}}\right)\hat{E}_{\mathrm{s}}^{(+)}\left(t'_{\mathrm{1}}\right)\rangle+\langle\hat{E}_{\mathrm{s}}^{(-)}\left(t_{\mathrm{2}}\right)\hat{E}_{\mathrm{s}}^{(-)}\left(t_{\mathrm{1}}\right)\hat{E}_{\mathrm{i}}^{(+)}\left(t'_{\mathrm{2}}\right)\hat{E}_{\mathrm{i}}^{(+)}\left(t'_{\mathrm{1}}\right)\rangle\nonumber\\
&&\hspace{10mm}+\langle\hat{E}_{\mathrm{s}}^{(-)}\left(t_{\mathrm{2}}\right)\hat{E}_{\mathrm{s}}^{(-)}\left(t_{\mathrm{1}}\right)\hat{E}_{\mathrm{s}}^{(+)}\left(t'_{\mathrm{2}}\right)\hat{E}_{\mathrm{i}}^{(+)}\left(t'_{\mathrm{1}}\right)\rangle+\langle\hat{E}_{\mathrm{s}}^{(-)}\left(t_{\mathrm{2}}\right)\hat{E}_{\mathrm{s}}^{(-)}\left(t_{\mathrm{1}}\right)\hat{E}_{\mathrm{i}}^{(+)}\left(t'_{\mathrm{2}}\right)\hat{E}_{\mathrm{s}}^{(+)}\left(t'_{\mathrm{1}}\right)\rangle\nonumber\\
&&\hspace{10mm}+\langle\hat{E}_{\mathrm{s}}^{(-)}\left(t_{\mathrm{2}}\right)\hat{E}_{\mathrm{i}}^{(-)}\left(t_{\mathrm{1}}\right)\hat{E}_{\mathrm{s}}^{(+)}\left(t'_{\mathrm{2}}\right)\hat{E}_{\mathrm{s}}^{(+)}\left(t'_{\mathrm{1}}\right)\rangle+\langle\hat{E}_{\mathrm{s}}^{(-)}\left(t_{\mathrm{2}}\right)\hat{E}_{\mathrm{i}}^{(-)}\left(t_{\mathrm{1}}\right)\hat{E}_{\mathrm{s}}^{(+)}\left(t'_{\mathrm{2}}\right)\hat{E}_{\mathrm{i}}^{(+)}\left(t'_{\mathrm{1}}\right)\rangle\nonumber\\
&&\hspace{10mm}+\left.\langle\hat{E}_{\mathrm{s}}^{(-)}\left(t_{\mathrm{2}}\right)\hat{E}_{\mathrm{i}}^{(-)}\left(t_{\mathrm{1}}\right)\hat{E}_{\mathrm{i}}^{(+)}
\left(t'_{\mathrm{2}}\right)\hat{E}_{\mathrm{s}}^{(+)}\left(t'_{\mathrm{1}}\right)\rangle+\langle\hat{E}_{\mathrm{s}}^{(-)}\left(t_{\mathrm{2}}\right)
\hat{E}_{\mathrm{i}}^{(-)}\left(t_{\mathrm{1}}\right)\hat{E}_{\mathrm{i}}^{(+)}\left(t'_{\mathrm{2}}\right)\hat{E}_{\mathrm{i}}^{(+)}\left(t'_{\mathrm{1}}\right)\rangle\right.\nonumber\\
&&\hspace{10mm}+\left\{{\rm s}\leftrightarrow {\rm i}\right\}\bigg]\nonumber\\
&=&I_{\mathrm{ssss}}+I_{\mathrm{iiii}}+I_{\mathrm{sisi}}+I_{\mathrm{siis}}+I_{\mathrm{isis}}+I_{\mathrm{issi}}.\hspace{7cm}
\label{Eq:Probability2}
\end{eqnarray}
\end{widetext}

Notice that we may write the TPA signal as a function of the
mutual delay between the beams given by
$\tau=\left(\tau_\mathrm{s}-\tau_\mathrm{i}\right)$. This delay
can easily be introduced in Eq.~(\ref{Eq:Probability1}) by making
the substitution
$\hat{E}_{\mathrm{s}}(t)\rightarrow\hat{E}_{\mathrm{s}}(t+\tau)$
for the signal field. The symbol $\{\mathrm{s\leftrightarrow i}\}$
stands for the contributions that are obtained by interchanging
the labels $\mathrm{s}$ and $\mathrm{i}$.

Interestingly, one can show that the TPA signal given by
Eq.~(\ref{Eq:Probability2}) contains only six non-vanishing
contributions, corresponding to the terms with even number of
indices $\mathrm{s,i}$. These are collected in the last line of
Eq.~(\ref{Eq:Probability2}). In the following we will discuss the
explicit form of each of these non-vanishing terms.

We start by describing the first two terms, these correspond to
the case where photons from a single signal (idler) beam are
absorbed. This means that the four-point correlation function
describing this process in the signal beam (and similarly in the
idler beam) is $\langle \hat{E}^{(-)}_{\mathrm{s}}
\hat{E}^{(-)}_{\mathrm{s}} \hat{E}^{(+)}_{\mathrm{s}}
\hat{E}^{(+)}_{\mathrm{s}}\rangle$, whose explicit form is given
by
\begin{widetext}
\begin{eqnarray}
I_{\mathrm{ssss}}&=&\frac{1}{\hbar^4}\int_{-\infty}^{\infty}dt_{\mathrm{2}}\int_{-\infty}^{t_{\mathrm{2}}}dt_{\mathrm{1}}\int_{-\infty}^{\infty}dt'_{\mathrm{2}}\int_{-\infty}^{t'_{\mathrm{2}}}dt'_{\mathrm{1}}M^{*}\left(t_{\mathrm{2}},t_{\mathrm{1}}\right)M\left(t'_{\mathrm{2}},t'_{\mathrm{1}}\right)\langle\hat{E}_{\mathrm{s}}^{(-)}\left(t_{\mathrm{2}}\right)\hat{E}_{\mathrm{s}}^{(-)}\left(t_{\mathrm{1}}\right)\hat{E}_{\mathrm{s}}^{(+)}\left(t'_{\mathrm{2}}\right)\hat{E}_{\mathrm{s}}^{(+)}\left(t'_{\mathrm{1}}\right)\rangle\nonumber\\
&=&\frac{1}{\hbar^2 4 \epsilon^2A^2c^2n_{\mathrm s}^2}\int_{-\infty}^{\infty}\
d\omega_{\mathrm
s}\int_{-\infty}^{\infty}d\omega'_{\mathrm
s}\mathcal{K}^{*}\left(\omega_{\mathrm
s}\right)\mathcal{K}\left(\omega'_{\mathrm
s}\right)\left[F_{1\mathrm s}\left(\varepsilon_{\mathrm f}-\varepsilon_{\mathrm g}-\omega_{\mathrm
 s},\omega'_{\mathrm s}\right)F_{1\mathrm s}\left(\omega_{\mathrm
s},\varepsilon_{\mathrm f}-\varepsilon_{\mathrm g}-\omega'_{\mathrm s}\right)\right.+\nonumber\\
&&\left.F_{1\mathrm s}\left(\varepsilon_{\mathrm f}-\varepsilon_{\mathrm g}-\omega_{\mathrm
s},\varepsilon_{\mathrm f}-\varepsilon_{\mathrm g}-\omega'_{\mathrm
s}\right)F_{1\mathrm s}\left(\omega_{\mathrm
s},\omega'_{\mathrm s}\right)\right],
\label{Eq:20}
\end{eqnarray}
\end{widetext}
where the spectral response of the medium is given by the function $\mathcal{K}\left(\omega\right)$,
\begin{equation}
\mathcal{K}\left(\omega\right)=\sum_k\frac{\mu_{\mathrm{f}k}\mu_{k\mathrm{g}}}{\varepsilon_{k}-\varepsilon_{\mathrm{g}}-\omega},
\end{equation}
and the spectral functions of the fields are described by $F_{\mathrm{1i,1s}}$. The spectral functions $F_\mathrm{1j}$, with
$\mathrm{j=i,s}$, contain information about the \emph{classical
correlations} of the photons and they are given by
\cite{torres-company2011}
\begin{equation}
F_{\mathrm{
1j}}\left(\omega,\omega'\right)=\sum_g\sqrt{\omega\omega'}f^*_{\mathrm{j},g}\left(\omega\right)f_{\mathrm{j},g}\left(\omega'\right)|v_{g}|^2.
\end{equation}
Note that, as expected, the first two non-vanishing terms of
Eq.~(\ref{Eq:Probability2}) do not depend on the mutual delay
between the signal and idler beams. This implies that these
contributions represent a background noise for VSS, as they do not
carry spectroscopic information about the sample.

The next contribution, $I_\mathrm{sisi}$, may explicitly be written as
\begin{widetext}
\begin{eqnarray}
I_{\mathrm{sisi}}\left(\tau_\mathrm{s},\tau_\mathrm{i}\right)&=&\frac{1}{\hbar^4}\int_{-\infty}^{\infty}dt_{\mathrm{2}}\int_{-\infty}^{t_{\mathrm{2}}}dt_{\mathrm{1}}\int_{-\infty}^{\infty}dt'_{\mathrm{2}}\int_{-\infty}^{t'_{\mathrm{2}}}dt'_{\mathrm{1}}M^{*}\left(t_{\mathrm{2}},t_{\mathrm{1}}\right)M\left(t'_{\mathrm{2}},t'_{\mathrm{1}}\right)\langle\hat{E}_{\mathrm{s}}^{(-)}\left(t_{\mathrm{2}}\right)\hat{E}_{\mathrm{i}}^{(-)}\left(t_{\mathrm{1}}\right)\hat{E}_{\mathrm{s}}^{(+)}\left(t'_{\mathrm{2}}\right)\hat{E}_{\mathrm{i}}^{(+)}\left(t'_{\mathrm{1}}\right)\rangle\nonumber\\
&=&\frac{1}{\hbar^2 4 \epsilon^2A^2c^2n_{\mathrm s}n_{\mathrm
i}}\int_{-\infty}^{\infty}\
d\omega_{\mathrm
i}\int_{-\infty}^{\infty}d\omega'_{\mathrm
i}\mathcal{K}^{*}\left(\omega_{\mathrm
i}\right)\mathcal{K}\left(\omega'_{\mathrm
i}\right)\exp\left[{i\left(\omega_{\mathrm
i}-\omega'_{\mathrm i}\right)
    \left(\tau_\mathrm{i}-\tau_\mathrm{s}\right)}\right]\nonumber\\
&&\hspace{-3mm} \times
 \left[F^*_2\left(\varepsilon_{\mathrm f}-\varepsilon_{\mathrm g}-\omega_{\mathrm
 i},\omega_{\mathrm
 i}\right)F_2\left(\varepsilon_{\mathrm f}-\varepsilon_{\mathrm g}-\omega'_{\mathrm
 i},\omega'_{\mathrm i}\right)+
   F_{1\mathrm i}\left(\omega_{\mathrm
   i},\omega'_{\mathrm i}\right)F_{1\mathrm s}\left(\varepsilon_{\mathrm f}-E_{\mathrm
   g}-\omega_{\mathrm
   i},\varepsilon_{\mathrm f}-\varepsilon_{\mathrm g}-\omega'_{\mathrm i}\right)\right].
   \label{Eq:23}
\end{eqnarray}
\end{widetext}
Notably, this term contains the function $F_2$, which is directly
related to the \emph{quantum correlations} between the fields and
is defined by \cite{torres-company2011}
\begin{equation}
F_{2}\left(\omega,\omega'\right)=\sum_g\sqrt{\omega\omega'}f_{\mathrm{s},g}\left(\omega\right)f_{\mathrm{i},g}\left(\omega'\right)v_{g}u_{g}.
\end{equation}
{\color{blue}}
The fourth term, $I_{\mathrm siis}$, is equal to the cross correlation
$\mathrm{(si)-(is)}$. Here, it is
important to highlight the fact that $ \mathrm{(si)}$ and $ \mathrm{(is)}$ are
not mutually interchangeable as time-ordering must be satisfied. Consequently,
this term writes
\begin{widetext}
\begin{eqnarray}
I_{\mathrm{siis}}\left(\tau_\mathrm{s},\tau_\mathrm{i}\right)&=&\frac{1}{\hbar^4}\int_{-\infty}^{\infty}dt_{\mathrm{2}}\int_{-\infty}^{t_{\mathrm{2}}}dt_{\mathrm{1}}\int_{-\infty}^{\infty}dt'_{\mathrm{2}}\int_{-\infty}^{t'_{\mathrm{2}}}dt'_{\mathrm{1}}
M^{*}\left(t_{\mathrm{2}},t_{\mathrm{1}}\right)M\left(t'_{\mathrm{2}},t'_{\mathrm{1}}\right)\langle\hat{E}_{\mathrm{s}}^{(-)}\left(t_{\mathrm{2}}\right)\hat{E}_{\mathrm{i}}^{(-)}\left(t_{\mathrm{1}}\right)\hat{E}_{\mathrm{i}}^{(+)}\left(t'_{\mathrm{2}}\right)\hat{E}_{\mathrm{s}}^{(+)}\left(t'_{\mathrm{1}}\right)\rangle\nonumber\\
&=&\frac{1}{\hbar^2 4 \epsilon^2A^2c^2n_{\mathrm s}n_{\mathrm
i}}\int_{-\infty}^{\infty}\
d\omega_{\mathrm
i}\int_{-\infty}^{\infty}d\omega_{\mathrm
s}\mathcal{K}^{*}\left(\omega_{\mathrm
i}\right)
\mathcal{K}\left(\omega_{\mathrm s}\right)\exp\left[{i\left(\omega_{\mathrm
i}+\omega_{\mathrm
s}\right)\left(\tau_\mathrm{i}-\tau_\mathrm{s}\right)+i\left(\varepsilon_{\mathrm f}-\varepsilon_{\mathrm g}\right)\left(\tau_\mathrm{s}-\tau_\mathrm{i}\right)}\right]\nonumber\\
&&\hspace{-3mm} \times \left[
F^{*}_{\mathrm {2}}\left(\varepsilon_{\mathrm f}-\varepsilon_{\mathrm g}-\omega_{\mathrm
i},\omega_{\mathrm
i}\right)F_{\mathrm {2}}\left(\omega_{\mathrm s},\varepsilon_{\mathrm f}-\varepsilon_{\mathrm g}-\omega_{\mathrm s}\right)+
F_{1\mathrm{s}}\left(\varepsilon_{\mathrm f}-\varepsilon_{\mathrm g}-\omega_{\mathrm
i},\omega_{\mathrm
s}\right)F_{1\mathrm{i}}\left(\omega_{\mathrm i},\varepsilon_{\mathrm f}-\varepsilon_{\mathrm g}-\omega_{\mathrm
s}\right)\right].
\label{Eq:24}
\end{eqnarray}
\end{widetext}
The remaining terms are easily obtained by interchanging the label
$\mathrm{s}$ and $\mathrm{i}$ in Eqs.~(\ref{Eq:23}) and
(\ref{Eq:24}), respectively.

\section{Virtual-state spectroscopy with entangled beams}

\begin{figure}[b!]
    \centering
    \includegraphics[width=8cm]{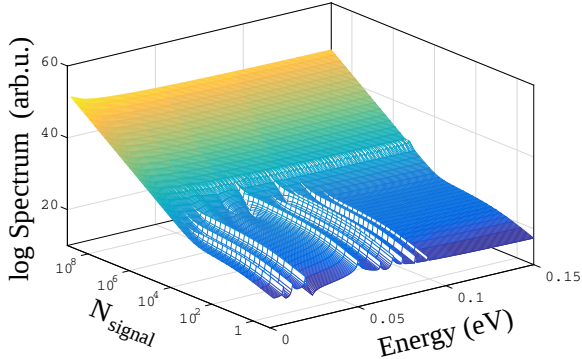}
    \caption{TPA spectrogram as a function of the number  $N_{signal}$ of photons in the
        twin beam; the delay range considered is $0\leq\tau\leq 8$ ps,
        the length of the nonlinear crystal is L = 1 mm and the pump pulse
        duration is set to $\tau_p$ = 1 ps, and the inverse group velocities
        are set to $G_s$ = 5.2 ps/m and $G_i$ = 5.6 ps/m. For the sake of simplicity, the energy axis is defined by means of the energy mismatch: $2\varepsilon_{k} - \varepsilon_{\mathrm{f}}$.}
    \label{Fig:spectrogram}
\end{figure}

We are now ready to discuss the implementation of the VSS protocol
with intense twin beams. For this, we consider a simple model
system in which the two-photon excitation energy of the medium
$|\mathrm{g}\rangle\rightarrow|\mathrm{f}\rangle$ corresponds to
the pump wavelength $\lambda_\mathrm{p0}=400$ nm
($E_{\mathrm{f}}=3.1$ eV considering $\varepsilon_{\mathrm{g}}=0$
eV). The intermediate-level energies are randomly chosen to be
$\varepsilon_{k}=\left(\varepsilon_{\mathrm{f}}+\{0.05, 0.075,
0.089\}\right)/2$ eV. Following this energy-level configuration,
we consider a nonlinear crystal producing degenerate photon pairs
with a wavelength of 800 nm. Notice that the central frequency of
the down-converted photons does not match any of the intermediate
states, thus making them effectively non-resonant (or virtual)
transitions. Moreover, the energies of these transitions, although
random, are kept close to the central frequency of the photons.
The reason for this lies in the fact that entangled two-photon
absorption is optimal under conditions of near-resonance between
entangled photons and the intermediate states \cite{upton2013}.
Finally, we assume the group velocity matching condition
\cite{keller1997}: $G_{\mathrm p}=(G_{\mathrm s}+G_{\mathrm
i})/2$, with $G_s$, $G_i$ and $G_p$ being in turn the inverse
group velocities of the signal, idler and pump beams. We have
considered this condition because it allows for a simple
interpretation and control of the correlations between the photon
beams.

\begin{figure}[t!]
    \begin{tabular}{cc}
        \includegraphics[width=8.5cm]{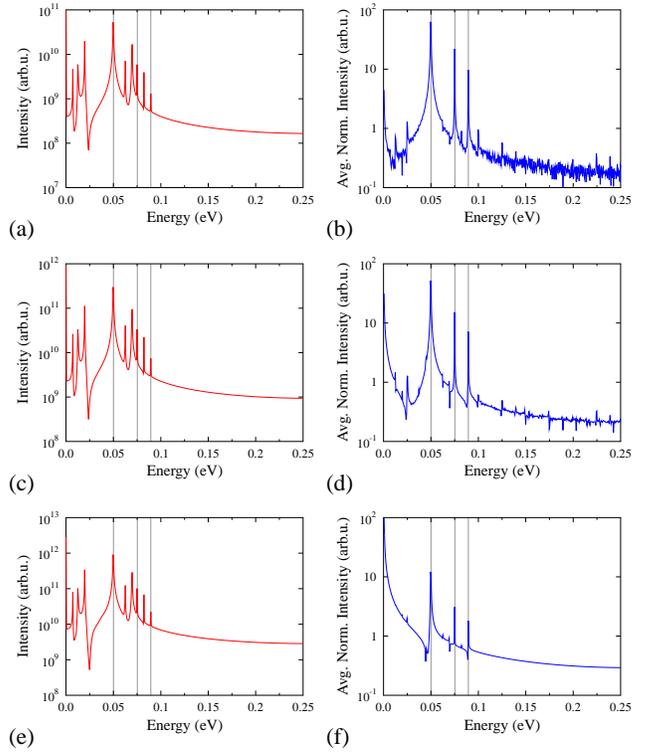}
    \end{tabular}
    \caption{TPA spectrograms, and their corresponding average, for twin beams
    carrying (a,b) $N_s=1$, (c,d) $N_s=10$, and (e,f) $N_s=100$ photons.
    Average was performed over and ensemble of 100 crystals of different
    lengths $L\in\llav{20,22}$ mm. The vertical grey lines indicate the
    relative energies, $2\varepsilon_{k} - \varepsilon_{\mathrm{f}}$, of the intermediate levels.}
    \label{Fig:average_spectrogram}
\end{figure}

Figure~\ref{Fig:spectrogram} shows the TPA spectrogram
--- Fourier transform w.r.t. the delay between beams, of the TPA
signal --- as a function of the number of photons
$N_\mathrm{signal}\equiv N_{\mathrm{s}}= N_{\mathrm{i}}$ carried
by the twin beam. We can immediately see from
Fig.~\ref{Fig:spectrogram} that several peaks emerge from the
Fourier transform of the TPA signal. These peaks appear as a
result of the interference between different pathways through
which two-photon excitation of the medium can occur
\cite{Saleh1998Entangled,Leon2013Role}. Concurrently, as pointed
out previously \cite{svozilik2016practical}, we can see that the
visibility of the TPA signal is affected as the number of photons
$N_{\text{signal}}$ is increased, thus defining a limit in the
photon flux at which VSS can be implemented. Interestingly, one
can make use of the signal shown in Fig.~\ref{Fig:spectrogram} to
extract information related to the energy level structure of the
medium
\cite{Saleh1998Entangled,Perina1998Multiphoton,Leon2013Role}. To
do so, we perform an average of the normalized TPA spectrograms
over different crystal lengths,
\begin{equation}
\tilde{S}_{g\rightarrow f}\pare{\tau} = \frac{1}{N}\sum_{n=1}^{N} \frac{S_{g\rightarrow f}\pare{\tau;L_n}}{\text{max}\llav{S_{g\rightarrow f}\pare{\tau;L_n}}},
\end{equation}
where $N$ is the number of crystal lengths used to obtain the
corresponding TPA signals. Notice that by changing the crystal
length, one effectively modifies the correlation (entanglement)
time between photons
\cite{Saleh1998Entangled,Perina1998Multiphoton,Leon2013Role},
which means that the average is effectively performed over
different correlation times. Therefore, we can experimentally
obtain the average of the TPA signals by using different
strategies, depending on the configuration of the setup that is
being used. For instance, in type-I SPDC, changing the width of
the pump beam modifies the correlation between photons
\cite{joobeur1994}, whereas in type-II, the configuration used in
this work, the correlation time is linearly proportional to the
crystal length \cite{shih1994}, so a proper set of two
wedge-shaped nonlinear/compensation crystals might be used.
Another alternative to control the correlation time of the photons
is by introducing frequency chirps in either the pump pulse or
down-converted beams \cite{svozilik2016practical}, or by changing
the time duration of the pump, as depicted in Fig.~4.

\begin{figure}[b!]
    \begin{tabular}{cc}
        \includegraphics[width=8.5cm]{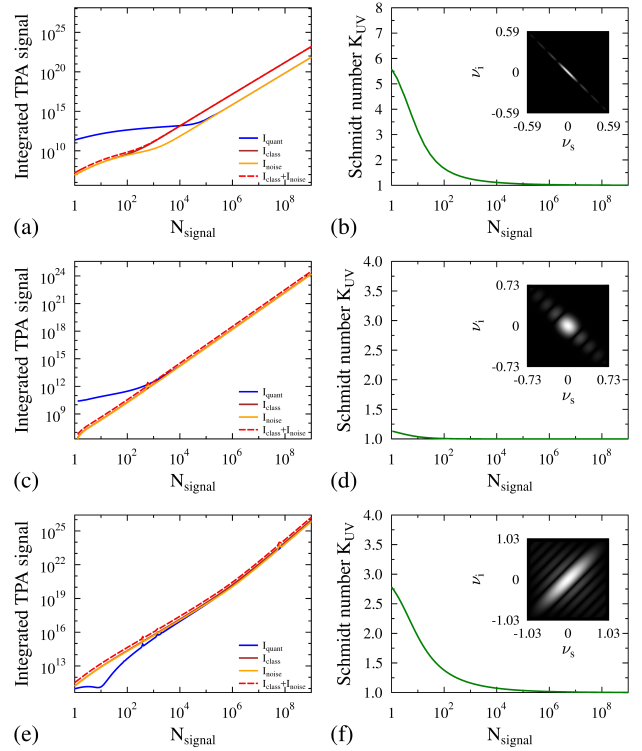}
    \end{tabular}
    \caption{Delay-integrated TPA signal (left column) for
    spectrally (a) anti-correlated,
    (c) uncorrelated, and (e) correlated two-photon states, and their
    corresponding Schmidt decomposition (right column). The
    symbols $\nu_{\rm s,i}=\omega_{\rm s,i} -
        \omega_{\rm s,i}^{0}$ stand for the frequency deviations from the
        central
        frequency $\omega_{s,i}^0$ of the photon fields. Integration of the
        signals was performed over a delay range $\Delta \tau$
        of $0 \leq\tau\leq 8$ ps. The length of the nonlinear crystal is
        $L=1$~mm, and the pump pulse duration was set to (a) $\tau_{\mathrm p}
        = 20$~fs, (c) $\tau_{\mathrm p} = 110$~fs, and (e) $\tau_{\mathrm p} =
        1$~ps.}
    \label{Fig:Threshold}
\end{figure}

Figure~\ref{Fig:average_spectrogram} shows the TPA spectrograms,
and their corresponding average over $100$ crystal lengths, for
twin beams carrying (a,b) $N_s=1$, (c,d) $N_s=10$, and (e,f)
$N_s=100$ photons. Notice that by averaging the TPA spectrograms
only three peaks remain in the signal, whose locations reveal the
energy of the intermediate states that contribute to the
two-photon excitation of the medium. Interestingly, the results in
Figs.~\ref{Fig:spectrogram} and \ref{Fig:average_spectrogram} show
that although VSS can be implemented with entangled fields
carrying a large number of photons, there exists a threshold in
the photon-pair number, as the height of the peaks with respect to
the background noise (signal-to-noise ratio) gets diminished with
increasingly larger photon fluxes. It is important to remark that
in obtaining the results shown in
Fig.~\ref{Fig:average_spectrogram}, we have assumed that the
temporal walk-off of the photons is much shorter than the lifetime
of the intermediate levels.

To explore the limits on the photon flux that can be used in the
implementation of VSS, we now analyze the terms that contribute to
the overall TPA signal. For the sake of simplicity, we divide all
terms into three different groups. The first group
($I_{\text{noise}}$) contains the terms that are independent of
the delay between the beams, namely the first two terms of
Eq.~(\ref{Eq:Probability2}). As we discussed above, these terms do
not contain spectroscopic information and represent a background
noise of the TPA signal. The second group ($I_{\text{class}}$)
contains the terms related to the classical correlations,
represented by those containing products of the functions
$F_\mathrm{1j}$, with $\mathrm{j=s,i}$ [see Eqs.~(\ref{Eq:23}) and
(\ref{Eq:24})]. Notice that these functions contain single-beam
correlations only. Finally, the third group ($I_{\text{quant}}$),
which depends on products of the functions $F_{2}$, represents the
quantum correlations between the two photon fields. This final
group is the most important one, as it contains the information
about the intermediate-state transitions occurring during the TPA
process.

Figure~\ref{Fig:Threshold} shows the delay-integrated TPA
contributions, considering three different initial states, namely
spectrally (a,b) anti-correlated, (c,d) quasi uncorrelated, and
(e,f) correlated photons. The delay-integrated signals are
obtained by integrating the TPA signal [Eq.
(\ref{Eq:Probability2})] over a delay interval $\Delta\tau$, while
fixing the correlation time between photons by means of the
pump-pulse duration $\tau_p$. Notice that the delay-integrated
signals with varying frequency correlations [depicted in
Figs.~\ref{Fig:Threshold}(b,d,f)] reflect the mutual interplay
between all participating terms (noise, classical, and quantum).
More importantly, they show that quantum contributions (blue solid
line), where spectroscopic information about the sample resides,
dominates in a low-to-moderate photon flux regime for
anti-correlated and quasi uncorrelated photons. This effect can be
understood in terms of the Schmidt modes of the two-photon state
[depicted in Figs.~\ref{Fig:Threshold}(b,d)]. As the number of
photons is increased, the number of effectively populated Schmidt
modes becomes smaller. The reduction of populated Schmidt modes
results in a reduced spectral overlap between the absorbed
entangled photons and the sample under study, thus lowering the
spectroscopic resolution of the system.

\begin{figure}[t]
    \begin{tabular}{cc}
        \includegraphics[width=8.5cm]{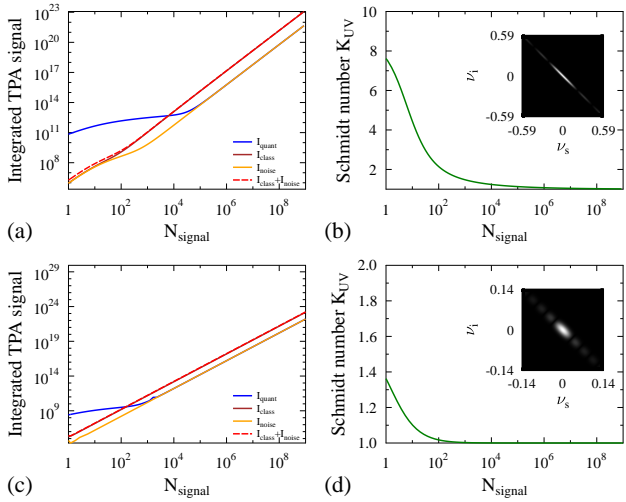}
    \end{tabular}
    \caption{Delay-integrated TPA signal {\color{blue}(left column)} for the
    fixed
        pump-duration $\tau_{\mathrm p}=1$~ps for different values of the
        crystal length (a)
        L=0.75~mm and (c) L=5~mm with the corresponding Schmidt
        numbers (right column) plotted in (b) and (d). Integration of the signals was
        performed over a delay range $\Delta \tau$} of $0
        \leq\tau\leq 8$ ps.
    \label{Fig:5}
\end{figure}

Remarkably, when correlated photons interact with the sample, we
find that the signal carrying useful information is completely
suppressed. This demonstrates the fact that entanglement alone is
not the key ingredient for implementing VSS, but rather a
combination of entanglement and anti-correlation of the absorbed
photons \cite{Leon2013Role}. In particular, for our model system,
the use of spectrally anti-correlated photons guarantees that the
transition from the quantum (linear) to the classical (quadratic)
regime occurs around $N_\mathrm{signal} \approx 10^4$ signal
photons. It is important to remark that this photon-number limit
is valid for the model system considered here. As we discussed
above, this limit strongly depends on the number of Schmidt modes
contained in the photon spectra. Indeed, as shown in
Fig.~\ref{Fig:5}, a broader spectrum of anti-correlated photons
will push the limit towards higher values of photon numbers,
whereas a narrower spectrum will reduce the photon flux at which
VSS can be successfully implemented.

Finally, notice that the Schmidt number $K_\mathrm{UV}$ is
inversely related to the length of crystal $L$, as depicted in
Fig.~\ref{Fig:5}. From here we can see that the use of shorter lengths of the
crystal produces a broader spectrum, with rich spatial structure, in the produced photon fields. This shows the many knobs that non-classical light provides to emerging nonlinear spectroscopy techniques.

\vspace{4mm}
\section{Conclusions}

We have presented a thorough analysis of the
virtual-state spectroscopy technique implemented with intense twin
beams. We showed that the virtual-state spectroscopy may be
implemented with entangled twin beams carrying up to $10^4$
photon pairs, provided that a proper configuration of the experimental setup
and a particular shape of the spectral correlations between photons is
selected. Our results suggest that by making use of intense twin beams one
might be able to detect two-photon absorption signals up to four orders of
magnitude larger than previously reported, thus paving the way towards the first
experimental implementation of the virtual-state spectroscopy
technique.

\section*{Acknowledgment}
JS thanks Alejandra Catalina Valencia Gonz\'{a}lez and Mayerlin Nu\~{n}ez
Portela for useful and stimulating discussions. JS also acknowledges the
Faculty of Science of Universidad de los Andes. This work was supported by
projects 17-23005Y (JS) and 15-08971S (JP) of the Czech Science
Foundation, and project LO1305 of M\v{S}MT \v{C}R. RJLM gratefully acknowledge financial support from DGAPA-UNAM, Mexico, under the project UNAM-PAPIIT IA100718.

\bibliography{VSS}
\end{document}